\documentclass[12pt,twocolumn]{emulateapj}
\usepackage{graphicx}
\usepackage{lipsum}
\usepackage{tabularx,amsmath,amssymb,color}
\usepackage{bm}
\usepackage{float}
\usepackage{natbib}
\newcommand{\beq}{\begin{equation}}
\newcommand{\eeq}{\end{equation}}

\newcommand{\ssc}{\scriptsize}
\newcommand{\ud}{\mathrm{d}}
\definecolor{dred}{rgb}{.9, .15,.1}
\definecolor{brown}{rgb}{0.5,0.2,0.2}
\definecolor{dgreen}{rgb}{0.15,0.6,0.15}

\def\half{{\textstyle{\frac{1}{2}}}}

\def\guide#1{\gdef\@guide{#1}}

\begin{document}

\bibliographystyle{apj}

\title{Adiabatic black hole growth in S\'ersic models of elliptical galaxies}

\author{Naveen Jingade\altaffilmark{1,2}, Tarun Deep Saini\altaffilmark{1}}

\author{Scott Tremaine\altaffilmark{3}}

\email{naveen@rri.res.in, tarun@physics.iisc.ernet.in}
\email{tremaine@ias.edu}

\altaffiltext{1}{Department of Physics, Indian Institute of Science,
  Bangalore 560 012, India}
\altaffiltext{2}{A \& A, Raman Research Institute, Bangalore 560080, India}
\altaffiltext{3}{Institute for Advanced Study, Princeton, NJ 08540, USA}

\begin{abstract}
\noindent
We have examined the effect of slow growth of a central black hole on
spherical galaxies that obey S\'ersic or $R^{1/m}$ surface-brightness
profiles. During such growth the actions of each stellar orbit are
conserved, which allows us to compute the final distribution function
if we assume that the initial distribution function is isotropic. We
find that black-hole growth leads to a central cusp or ``excess
light'', in which the surface brightness varies with radius as
$R^{-1.3}$ (with a weak dependence on S\'ersic index $m$), the
line-of-sight velocity dispersion varies as $R^{-1/2}$, and the
velocity anisotropy is $\beta\simeq -0.24$ to $-0.28$ depending on
$m$. The excess stellar mass in the cusp scales approximately linearly
with the black-hole mass, and is typically 0.5--0.85 times the
black-hole mass. This process may strongly influence the structure of
nuclear star clusters if they contain black holes.

\end{abstract}
 
\maketitle

\section{Introduction}

\noindent 
Most elliptical galaxies have surface-brightness profiles that are fit
well by the empirical S\'ersic formula, given by equation
(\ref{sersic}), with S\'ersic index $m$ varying from $2$ to much
larger values \citep{KOR09}. In addition to the index $m$,
the S\'ersic formula has just two parameters: $I_0$ is the central surface
brightness and $R_e$ is the effective radius, the projected radius
containing half of the total luminosity.

\begin{figure}[h]
\includegraphics[width=0.45\textwidth]{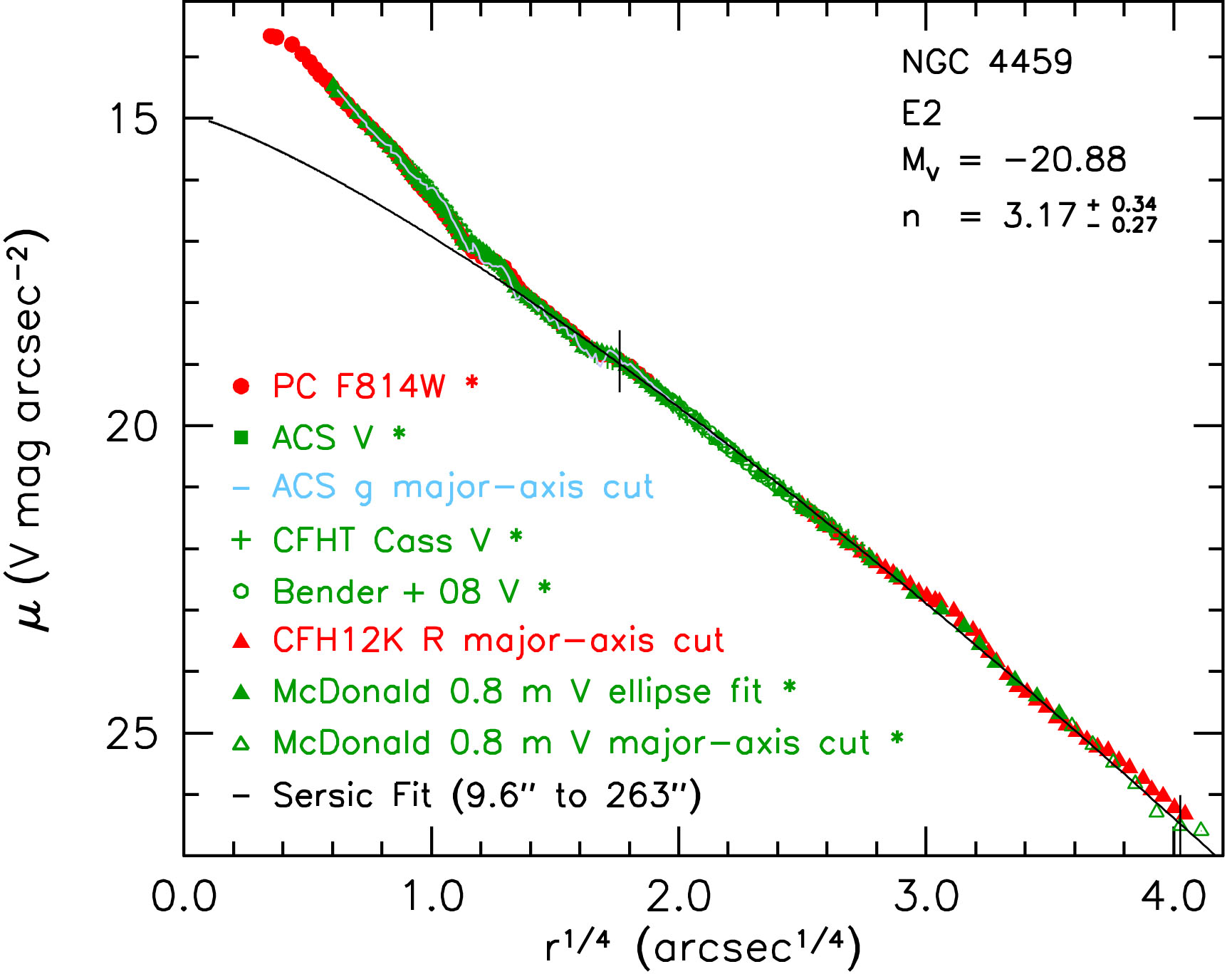}
\caption{\small Surface-brightness profile of the dwarf galaxy
  NGC 4459, an excess-light elliptical. The curve shows the
  best-fitting S\'ersic profile, with index $m=3.17$
  \citep{KOR09}. Reproduced by permission of the AAS.}
\label{fig:1}
\end{figure}

\begin{figure}[h]
\includegraphics[width=0.45\textwidth]{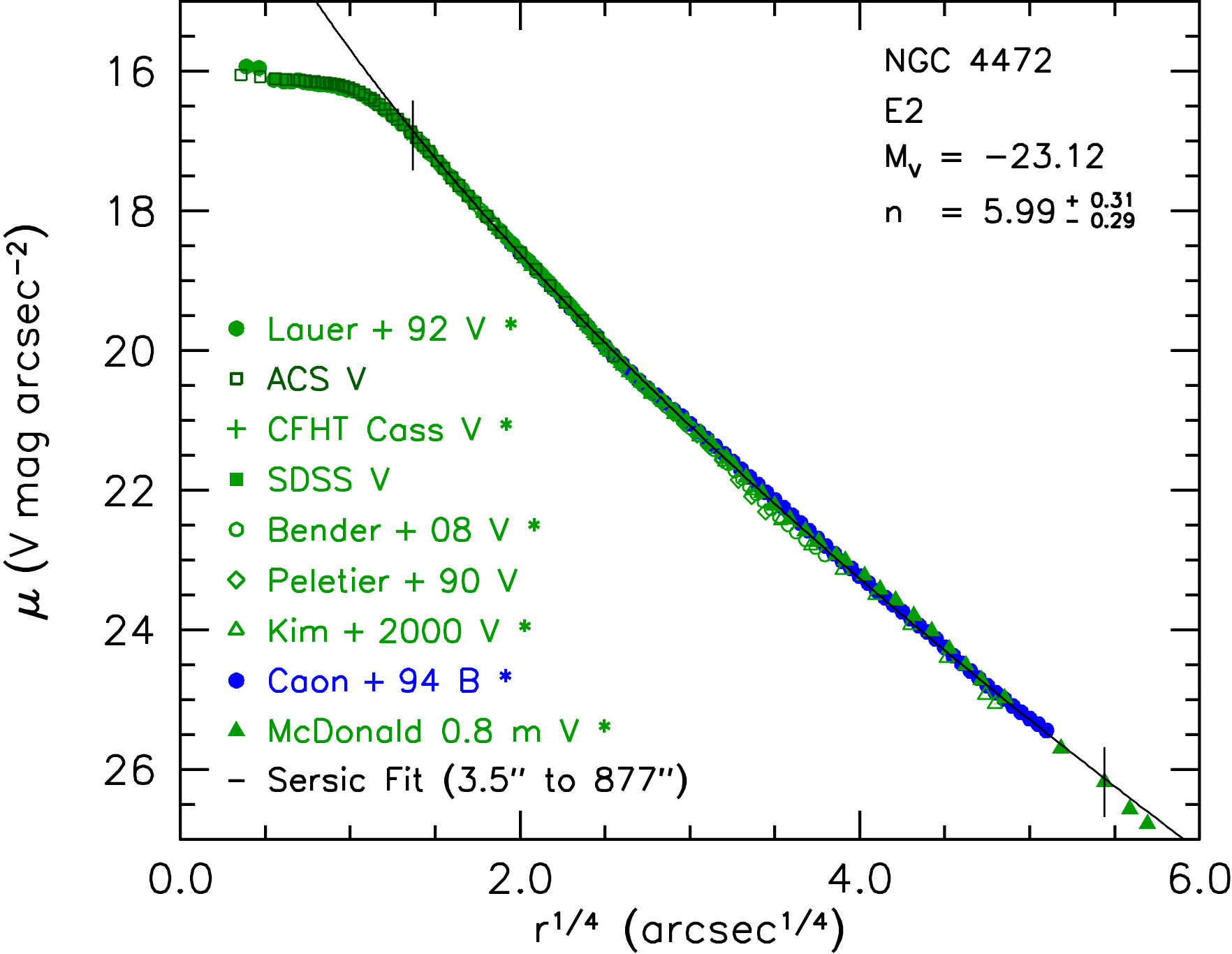}
\caption{\small Surface-brightness profile of the giant galaxy
  NGC 4472, a missing-light elliptical. The curve shows the
  best-fitting S\'ersic profile, with index $m=5.99$ \citep{KOR09}. Reproduced by permission of the AAS.}
\label{fig:2}
\end{figure}

It has been observed \citep{KOR09} that the S\'ersic profile fits
elliptical galaxies over a wide range of radii, but there are often
deviations from the fitting formula at small radii. These deviations
are of two kinds: either ``missing light'' or ``excess light'' near
the centers compared to the inward extrapolation of the S\'ersic
profile. Missing-light or ``core'' ellipticals generally have higher
luminosities ($M_V \lesssim -21.7$) than excess-light ellipticals
($M_V \gtrsim -21.5$). Figures \ref{fig:1} and \ref{fig:2}, taken from \cite{KOR09},
show the surface-brightness profiles of an excess-light and a
missing-light elliptical galaxy. 

The formation of a core is generally attributed to the dynamical
excitation of stellar orbits due to a binary black hole (hereafter BH), although
there is little or no direct evidence for this hypothesis. Excess light is
generally attributed to late star formation from gas that has become
concentrated near the galaxy center. The division of ellipticals into
these two classes, and the explanations given above for this
dichotomy, date back at least to \cite{fab97}.

Most elliptical galaxies appear to have massive BHs at their
centers. In this work we assume that these BHs form slowly by the
accretion of gas from larger radii. In this case the surrounding
stellar system will contract as the central mass grows, because the stellar
orbits conserve their adiabatic invariants or actions. If the galaxy
began with a S\'ersic surface-brightness profile, this process will
naturally produce excess light relative to the S\'ersic profile near
the center of the galaxy, and the corresponding excess stellar mass
will be simply related to the mass of the BH and the properties of the
initial S\'ersic profile. The main goal of the paper is to work out
these relations and to compare the predictions of this simple model to
the observations.

Pioneering studies of this process were carried out by \cite{PEB72}
and \cite{Y80}. These studies assumed that the galaxies were
spherical, as do we, but in contrast they assumed that the stellar
distribution function (DF) near the center of the galaxy was
Maxwellian before the formation of the BH. They found that as the BH
grew the stellar DF developed a cusp in which the density varied as
$\rho\sim r^{-3/2}$, corresponding to a surface brightness $I(R)\sim
R^{-1/2}$. However, a S\'ersic profile does not have a Maxwellian DF,
and the properties of the cusp formed by the adiabatic growth of a
central BH depend strongly on the DF \citep{qhs95}. In general we
expect (and shall find) the cusp formed in an initial S\'ersic model
to be steeper than the one found by Peebles and Young.

In \S\ref{sec:setup} we review the properties of S\'ersic models and
derive the dynamical quantities of interest (density, potential, DF,
velocity moments) assuming spherical symmetry, constant mass-to-light
ratio $\Upsilon$, and isotropic velocity dispersion. We then slowly
add a central point mass representing the BH and evolve the system
under the assumption of adiabatic invariance of the stellar
orbits. The numerical methods are described in \S\ref{sec:num}. The
surface-brightness distribution, anisotropy parameter, excess mass,
and line-of-sight velocity-dispersion distribution of the adiabatic
models are described in \S\S \ref{sec:excess} and \ref{sec:discuss},
and the observations in \S\ref{sec:obs}.  The conclusions are in
\S\ref{sec:conc}.

\section{Initial set up}

\label{sec:setup}

\subsection{Density}

\noindent
The surface-brightness distribution of a spherical galaxy is modeled
well by the S\'ersic law \citep{Sersic63,Sersic68,CIO91}
\begin{equation}
I(\eta) = I_0\,\exp(-b\,\eta^{1/m}) \quad \text{where}\quad \eta = R/R_e
\label{sersic}
\end{equation}
and $m$ is the S\'ersic index. Here $R_e$ is the effective radius, the radius on the sky that contains
half the total light. This definition of $R_e$ requires that the
parameter $b$ is given by 
\begin{equation}
2\,\gamma(2m,b)=\Gamma(2m)
\label{gamma}
\end{equation} 
where $\gamma$ is the incomplete gamma function.  The function
$b(m)$ is well fitted by the linear interpolation $b(m)=2m-0.324$,
with relative error less than $0.001$ for $0.5<m<10$
\citep{CIO91}. The exact values of $b(m)$ for several values of $m$
are given in Table \ref{tab1}. As $m\to\infty$, $b/m\to 2$ and
$I(R)\sim R^{-2}$ over a wide range of radii around $R_e$.

The luminosity density for given surface brightness $I(R)$ can be
obtained by the Abel integral 
\begin{equation}
j(r) = -\frac{1}{\pi}\,\int_r^\infty \frac{\mathrm{d}R}{\sqrt{R^2-r^2}}\frac{\mathrm{d}I}{\mathrm{d}R}.
\label{abel}
\end{equation} 

\begin{figure}[h]
\hspace{-1em}
\includegraphics[width=0.47\textwidth]{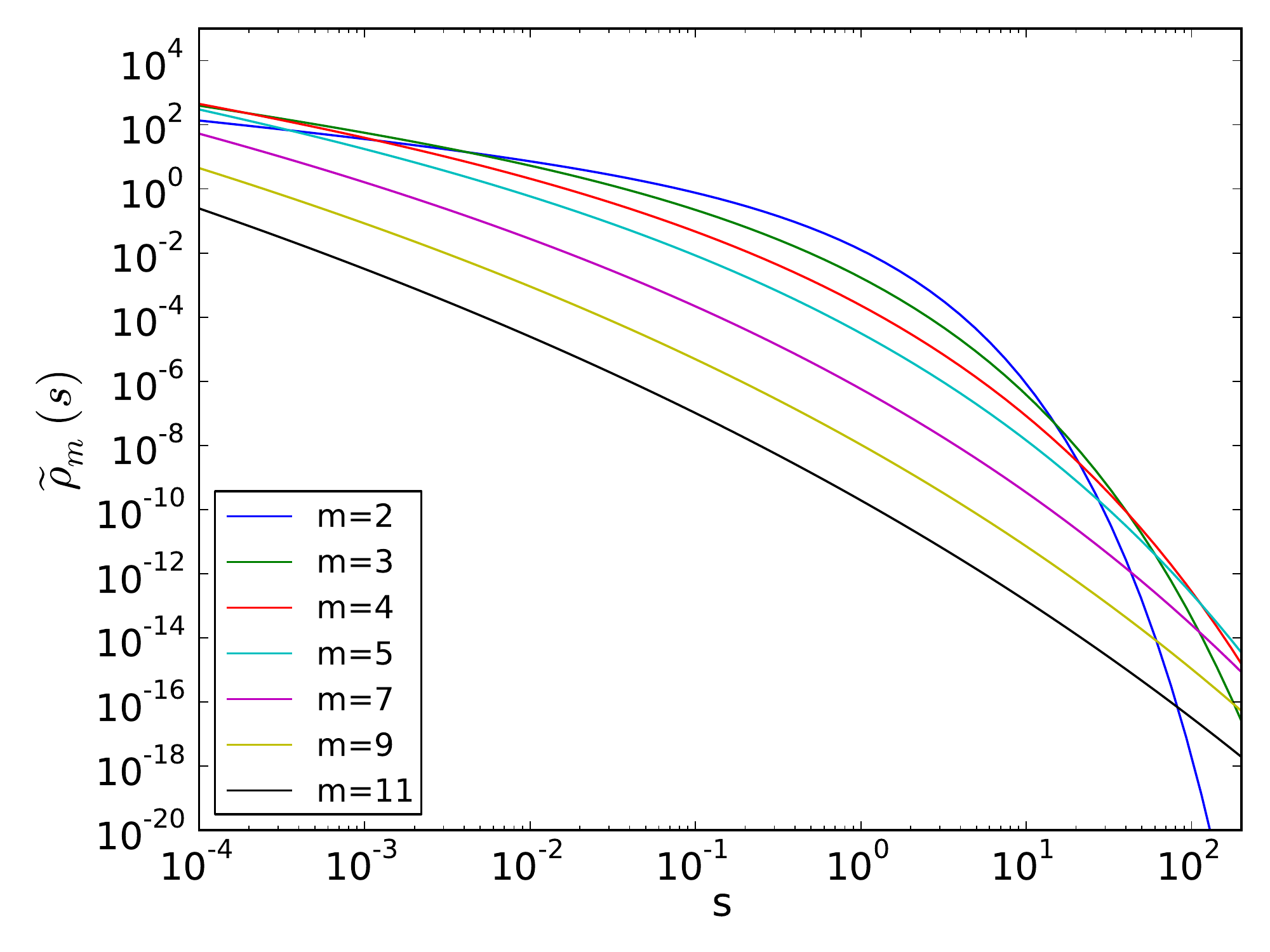}
\caption{The dimensionless density $\widetilde\rho_m(s)$ (eq.\
  \ref{rho}) for $2\leq m\leq 11$. Here $s=r/R_e$, the ratio of the
  radius to the effective radius.}
\label{fig:3}
\end{figure}

If we assume that the galaxy has constant mass-to-light ratio
$\Upsilon$, then its mass density $\rho(r)=\Upsilon j(r)$.
Substituting equation (\ref{sersic}) in equation (\ref{abel}) and
introducing the dimensionless radius $s=r/R_e$, we get
\begin{eqnarray}
\rho_m(s) & = &\Upsilon \frac{I_0}{R_e} \widetilde{\rho}_m(s)= \frac{\Sigma_0}{R_e} \widetilde{\rho}_m(s)\quad\mbox{where}\quad \Sigma_0 = \Upsilon\,I_0,\nonumber \\
\widetilde{\rho}_m(s)& = &\frac{b}{\pi\,ms^{(m-1)/m}}\int^1_0
\frac{\exp[-b(s/x)^{1/m}]}{x^{1/m}}\frac{\ud x}{\sqrt{1-x^2}}  \quad
\label{rho}
\end{eqnarray} 

where $x = s/\eta$. For $m>1$ the dimensionless density $\widetilde{\rho}_m(s)$ diverges at the origin as
\begin{equation} 
\widetilde{\rho}_m(s)\sim\frac{b}{2\pi m}\,\frac{1}{s^{(m-1)/m}}B\big[\half,\half(1-1/m)\big]
\label{scaling}
\end{equation}
where $B$ is the beta function. The dimensionless density
$\widetilde{\rho}_m(s)$ is plotted in Figure \ref{fig:3} for several
values of $m$.

\subsection{Potential}

\noindent
The potential is obtained by solving the Poisson equation
\begin{equation}
\nabla^2 \Phi=\frac{1}{r^2}\frac{\ud}{\ud r}\left(r^2\frac{\ud
    \Phi}{\ud r} \right)=4\pi G \rho.
\end{equation}
If we define $\Phi_m(r) = 4\pi G\, \Sigma_0 R_e
\widetilde{\Phi}_m(s)$, then in dimensionless variables we have 
\begin{equation}
\frac{1}{s^2}\frac{\ud}{\ud s}\left(s^2\frac{\ud \widetilde{\Phi}_m}{\ud s} \right)=\widetilde{\rho}_m.
\label{poisson}
\end{equation}
In integral form,
\begin{eqnarray}
\widetilde{\Phi}_m(s)&=&-\int^\infty_s \frac{\widetilde{M}_m(s')}{s'^2}\ud
s'\quad\mbox{where}\nonumber \\
 \widetilde{M}_m(s)&=&\int_0^s\widetilde\rho_m(s')s'^2\,\ud s'
\label{eq:potnorm}
\end{eqnarray}

is the mass of the system measured in units of $4\pi\,\Sigma_0
R^2_e$. One cannot solve the integral in general for the density given
by equation (\ref{rho}) but we can obtain the central potential
\citep{CIO91}, which is useful in what follows: 
\begin{equation}
\Phi(0)=-4G\Upsilon\int_0^\infty \ud R \,I(R)=-4\pi G \int^\infty_0 r\,\rho(r)\,\ud r
\end{equation}
from which the dimensionless central potential is found to be:
\begin{equation}
\widetilde{\Phi}_m(0)=-\frac{\Gamma(m+1)}{\pi b^m}.
\label{inc}
\end{equation}
Thus the central potential is finite even
though the density diverges at the center for $m>1$.  The values of the
dimensionless central potential and dimensionless mass for several
values of the S\'ersic index $m$ are given in Table \ref{tab1}.

\begin{figure}[h]
\hspace*{-1em}
\includegraphics[width=0.48\textwidth]{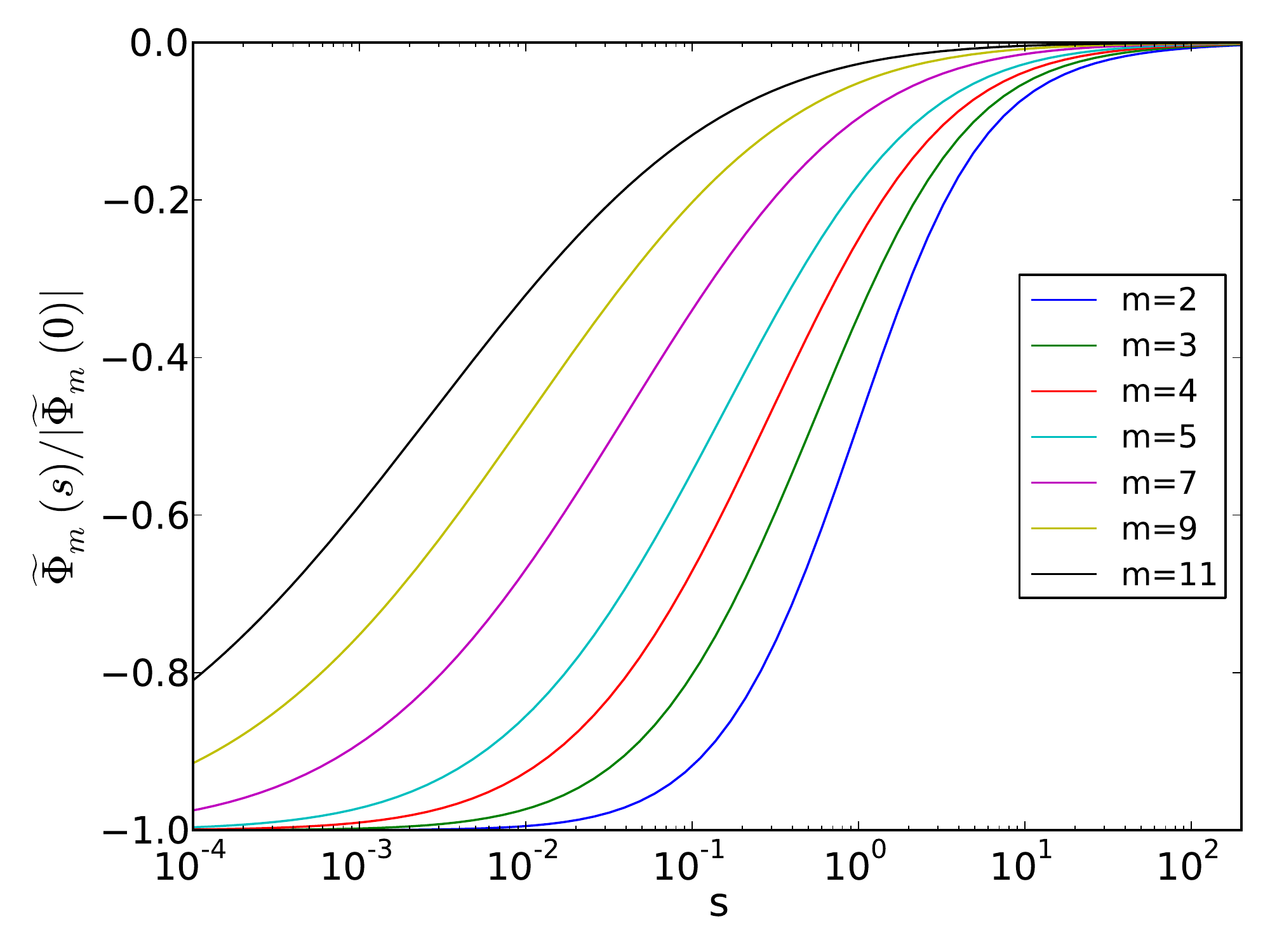}
\caption{The normalized potential
  $\widetilde{\Phi}_m(s)/|\widetilde{\Phi}_m(0)|$ (eqs.\
  \ref{eq:potnorm} and \ref{inc}) for $2\leq m\leq 11$. Here
  $s=r/R_e$, the ratio of the radius to the effective radius.}
\label{fig:4}
\end{figure}

Poisson's equation (\ref{poisson}) has to be solved numerically with
the initial condition (\ref{inc}), taking $\ud\widetilde{\Phi}_m/\ud
s=0$ at $s=0$.  Plots of the potential for different values of $m$
are shown in Figure \ref{fig:4}. 

\begin{table}[h]
 \setlength{\tabcolsep}{12pt}
 \centering
\caption{The dimensionless parameters $b$, $|\widetilde{\Phi}_m(0)|$
  and $\widetilde{M}_m$ for different values
  of the S\'ersic index $m$.}
\label{tab1}
\begin{tabular}{ccccc}
\hline\hline
\rule{0pt}{3.1ex} 
\rule[-1.2ex]{0pt}{0pt}
$m$ &
$b$ &
$|\widetilde{\Phi}_m(0)|$ &
$\widetilde{M}_m$ \\
\hline
\rule{0pt}{3.1ex} 
\rule[-1.2ex]{0pt}{0pt}
\hspace*{-1em}
2 & 3.67 \qquad & 4.17$\times 10^{-2}$\qquad & 2.91$\times 10^{-2}$\\
3 & 5.67 \qquad & 1.04$\times 10^{-2}$\qquad & 5.41$\times 10^{-3}$ \\
4 & 7.66 \qquad & 2.21$\times 10^{-3}$\qquad & 8.42$\times 10^{-4}$ \\
5 & 9.66 \qquad & 4.52$\times 10^{-4}$\qquad & 1.27$\times 10^{-4}$ \\
7 & 13.66\qquad & 1.80$\times 10^{-5}$\qquad & 2.74$\times 10^{-5}$ \\
9 & 17.66\qquad & 6.88$\times 10^{-7}$\qquad & 5.68$\times 10^{-8}$ \\
11 & 21.66 \qquad & 2.57$\times 10^{-8}$\qquad & 1.17$\times 10^{-9}$ \\
\hline
\end{tabular}
\end{table}

\subsection{Distribution function}

\begin{figure}[h]
\hspace*{-1em}
\includegraphics[width=0.48\textwidth]{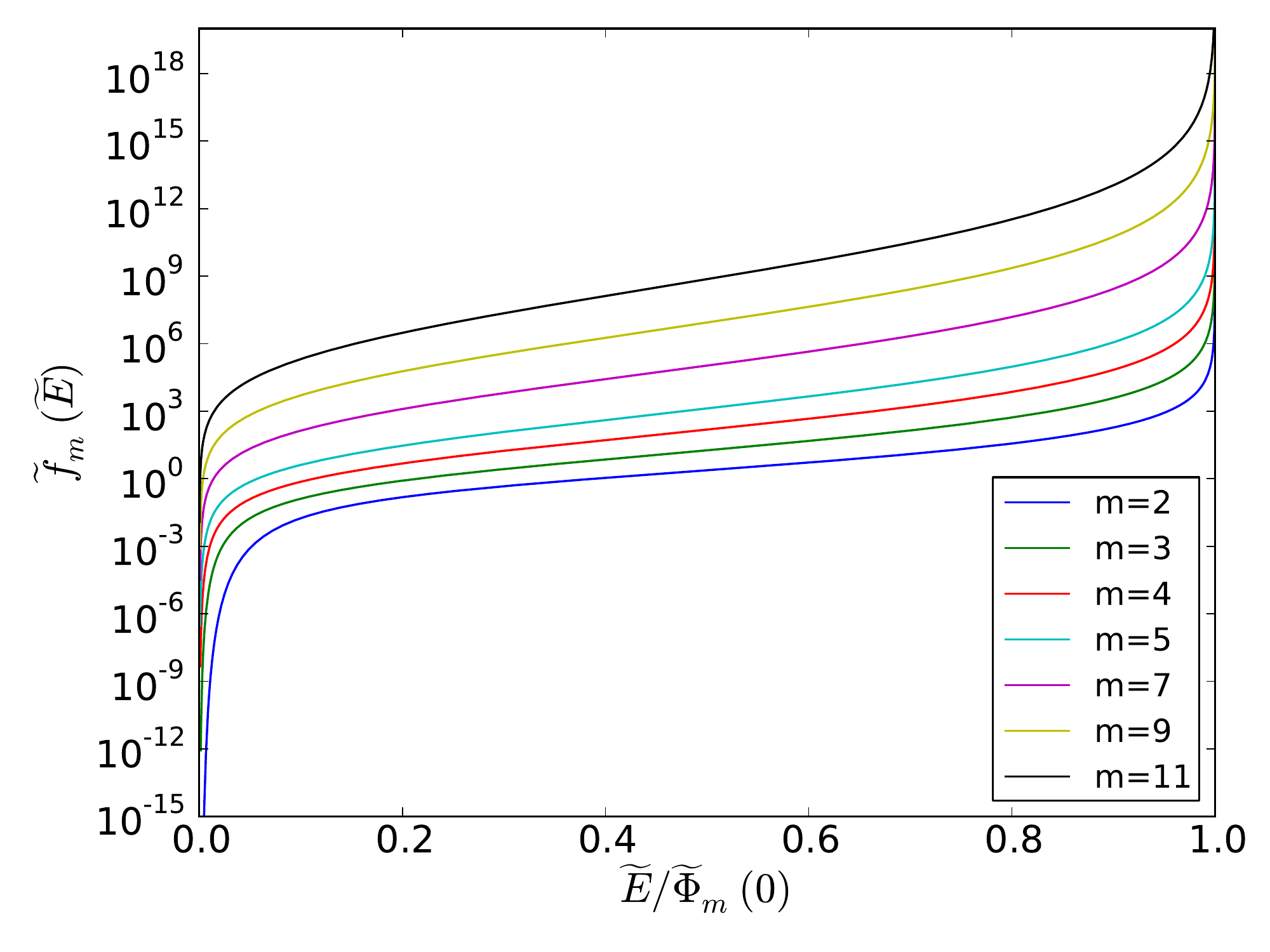}
\caption{The DF $\widetilde{f}_m({\widetilde{E}})$ as a function of
  normalized energy $\widetilde{{E}}/\widetilde{\Phi}_m(0)$ for $2\leq
  m\leq 11$.}
\label{fig:5}
\end{figure}

\noindent
We shall assume for simplicity that the initial phase-space DF of the stars in the galaxy is ergodic, that
is, the velocity distribution is isotropic at every point in space and
the DF depends only on the energy $E$. In such a case, the DF can be
calculated from the known density and potential of the galaxy as follows:
\begin{equation}
f({E})=\frac{1}{\sqrt{8\,}\pi^2}\int^0_{E} \frac{\ud \Phi}{\sqrt{\smash[b]{\Phi-E}}}\:\frac{\ud^2\rho}{\ud \Phi^2}.
\end{equation}
Using the dimensionless potential and density, we can write the DF in dimensionless form as
\begin{eqnarray}
f_m({E})&=&\frac{1}{(4\pi G)^{3/2}({\Sigma_0}
R_e^5)^{1/2}}\widetilde{f}_m({\widetilde{E}})\qquad\mbox{where}\nonumber\\
\widetilde{f}_m({\widetilde{E}})&=&\frac{1}{\sqrt{8\,}\pi^2} \int^0_{\widetilde{{E}}}
\frac{\ud
  \widetilde{\Phi}}{\sqrt{\smash[b]{\widetilde{\Phi}-\widetilde{{E}}}}}\:\frac{\ud^2\widetilde{\rho}_m}{\ud
  \widetilde{\Phi}^2}.
  \label{eq:edd}
\end{eqnarray}
The DF is plotted for various values of $m$ in Figure
~\ref{fig:5}. One can see that for all values of $m$ the DF is
positive, and that it diverges as ${E}\to\Phi(0)$; it is
straightforward to show that 
\begin{equation}
\widetilde{f}_m({\widetilde{E}})\sim \left[\widetilde{E}-\widetilde{\Phi}_m(0)\right]^{-\alpha},\qquad \alpha =\frac{5m+1}{2(m+1)}.
\label{eq:fdiv}
\end{equation}

\subsection{Velocity moments}

\noindent
For the remainder of this paper we will drop the tildes on $E$,
$\Phi$, $\rho$, etc., and it should be understood that these
quantities are dimensionless. According to Jeans's theorem, a stationary
spherical DF can only depend on the energy $E=\half v_r^2+\half
v_t^2+\Phi(s)$ and the angular momentum $L=s\,v_t$, where $v_r$ and
$v_t$ are the radial and tangential velocities. We can obtain the
density, and the radial and tangential velocity dispersions for this DF
as

\begin{widetext}
\begin{align}
  \rho(s)&=4\pi\int_{\Phi(s)}^0 \ud
  E\,[2(E-\Phi)]^{1/2}\int_0^1\frac{\ud y\,y\,f(E,L_{\rm
      max}\,y)}{\sqrt{1-y^2}},\label{rhoel} \\
 \overline{v_r^2}(s)&=\frac{4\pi}{\rho(s)}\int_{\Phi(s)}^0\ud
 E\,[2(E-\Phi)]^{3/2}\int_0^{1}\ud y \,y\,\sqrt{1-y^2}\,f(E,L_{\rm max}\,y),
 \label{vr}\\
 \overline{v_t^2}(s)&=\frac{4\pi}{\rho(s)}\int_{\Phi(s)}^0\ud
 E\,[2(E-\Phi)]^{3/2}\int_0^{1}\frac{\ud y \,y^3\,f(E,L_{\rm max}\,y)}{\sqrt{1-y^2}},
 \label{vt}
\end{align}
\end{widetext}

where 
\begin{equation} 
L_{\rm max}(E,s)=\{2\,s^2[E-\Phi(s)]\}^{1/2},\quad y=L/L_{\rm max} ;
\end{equation}
$L_{\rm max}$ is the maximum angular momentum for an orbit of energy
$E$ that passes through radius $s$. Note that the velocity dispersions
in the $\theta$ and $\phi$ directions are
$\overline{v_\phi^2}=\overline{v_\theta^2}=\half\overline{v_t^2}$. For
numerical work, the inner integral over $y$ can be evaluated using
Gauss--Chebyshev quadrature and the outer integral over $E$ by
Gauss--Legendre quadrature.

We then project the density and velocity moments on the plane of sky to obtain the surface density
\beq
\Sigma(\eta) = 2\int^{\infty}_\eta \frac{s\,\rho(s)\;\ud s}{\sqrt{s^2-\eta^2}}
\eeq
and the line-of-sight velocity dispersion
\beq
\sigma^2_{||}(\eta)=\frac{2}{\Sigma(\eta)}\int^{\infty}_\eta
\frac{s\,\rho(s)\;\ud
  s}{\sqrt{s^2-\eta^2}}\left[\left(1-\frac{\eta^2}{s^2}
  \right)\overline{v_r^2}
  +\frac{\eta^2}{2\,s^2}\;\overline{v_t^2}\right]. 
\label{eq:los}
\eeq

The anisotropy parameter is 
 \begin{equation}
 \beta(s) = 1-\frac{ \overline{v_t^2}(s)}{ 2\,\overline{v_r^2}(s)}
\label{eq:betadef}
 \end{equation}
which is $0$ for an isotropic velocity distribution, $-\infty$ for circular orbits and $+1$ for
radial orbits. 

\begin{figure}[h]
\hspace*{-1em}
\includegraphics[width=0.48\textwidth]{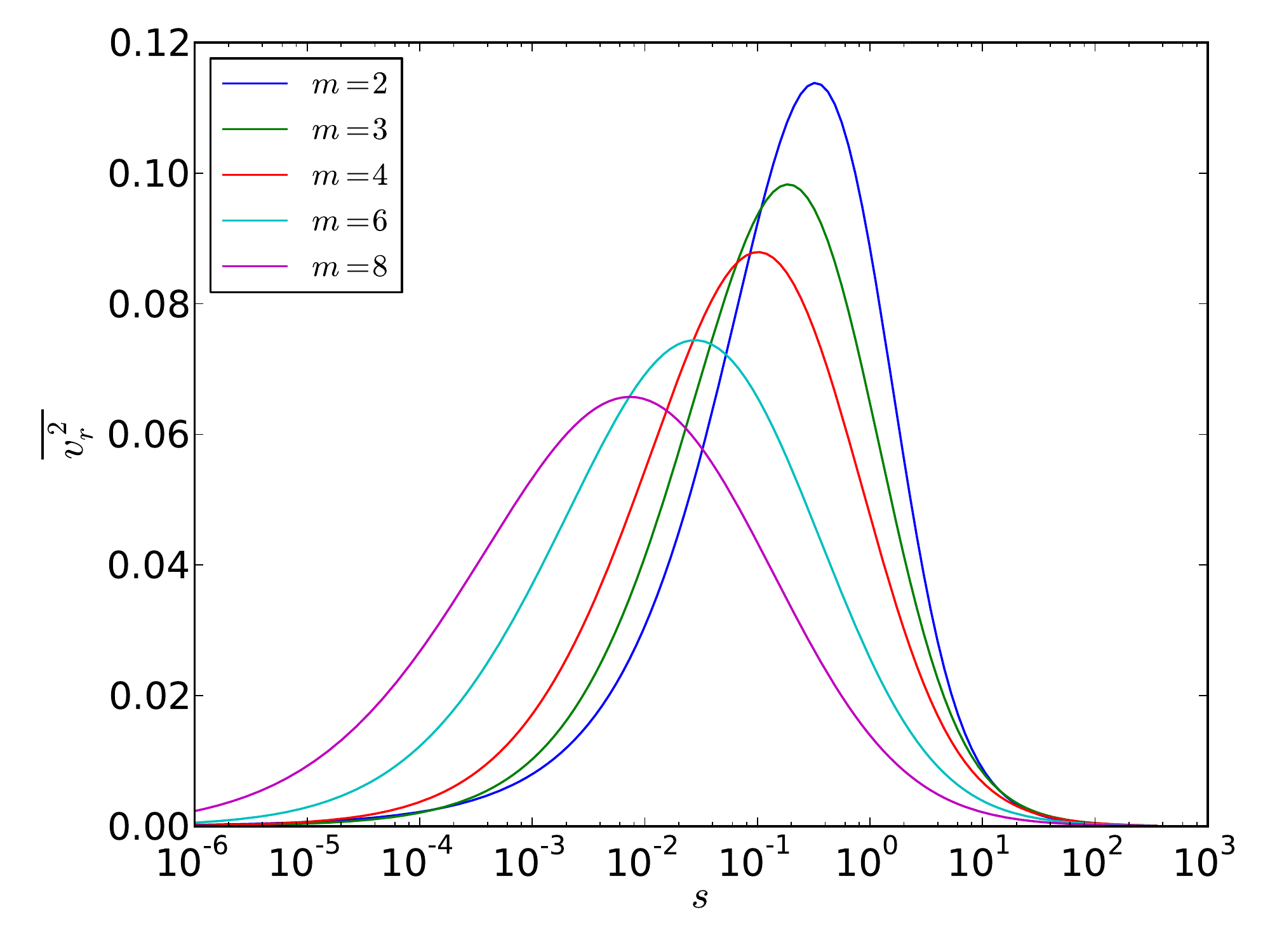}
\caption{\small The initial isotropic velocity dispersion of the galaxy for varying S\'ersic index $m$.}
\label{fig:14}
\end{figure}

The initial S\'ersic models are ergodic, so the velocity distribution
is isotropic, $\beta=0$, and the DF depends only on energy, not angular
momentum. The velocity dispersion
$\overline{v_r^2}(s)=\half\overline{v_t^2}(s)$ is shown in Figure \ref{fig:14}.

\subsection{Actions}
 
\noindent
The radial and azimuthal actions are 
\begin{equation}
 I_r = 2\int_{s_-}^{s_+}\sqrt{2[E-\Phi(s)]-L^2/s^2}\;\ud s \qquad
 I_t = 2\,\pi\, L,
\label{eq:action}
\end{equation}  
where $s_-$ and $s_+$ are the pericenter and apocenter distances of
the orbit, given implicitly by 
\begin{equation}
2\,s_{\pm}^2[E-\Phi(s_{\pm})]-L^2=0.
\end{equation}
Since actions are conserved during the adiabatic growth of the central
BH, the DF is also conserved at each point in action space. In
terms of energy and angular momentum, a star that is initially at
$(E,L)$ in phase space when the potential is $\Phi(s)$ will move to a
new phase point $(E^*,L)$ when the potential changes to
$\Phi^*(s)$. By equating the initial and final actions, we find the
relation $E=E(E^*,L)$. Then the new DF $f^*(E^*,L)$ is related to the
original DF $f(E,L)$ by $f^*(E^*,L)=f[E(E^*,L),L]$.

\section{Numerical methods}

\label{sec:num}

\noindent
The density, DF, potential and other galaxy properties are evaluated
on a grid: $\rho_i=\rho(s_i)$, $\Phi_i=\Phi(s_i)$,
$f_{jk}=f(E_j,L_k)$, and so on. The radial grid is equally spaced in
$\log (s)$ with roughly $200$ points between $s_{\rm min}=10^{-6}$ and
$s_{\rm max} =2\times10^2$. The energy grid points $E_j$ are chosen to
match the potential $\Phi_i$ at the radial grid points; since
$\Phi(r)$ varies as the BH grows, the locations of the energy
grid points $E_j$ also vary.  The grid points for angular momentum are
spaced uniformly between $x=0$ and $x=1$ where $x=L/L_c(E)$ and $L_c(E)$
is the angular momentum of a circular orbit at the given
energy. Typically we use 200 grid points
in energy and 20 in angular momentum.

Following \cite{Y80}, the program executes the following steps for a
given S\'ersic index $m$:

\begin{enumerate}

\item The density $\rho_i=\rho(s_i)$ is obtained from equation
  (\ref{rho}) at all points on the radial grid (recall that we have
  dropped the tildes on all dimensionless quantities).

\item The gravitational potential $\Phi_i=\Phi(s_i)$ is obtained from
  Poisson's equation (\ref{eq:potnorm}) at all points on the grid. 

\item The initial DF $f_{jk}=f(E_j)$ is obtained from $\rho_i$ and $\Phi_i$
  using equation (\ref{eq:edd}). The initial values $f_{jk}$ are
  independent of the angular momentum $x_k\,L_c(E_j)$ since the initial
  DF is ergodic. 

\item The radial action $I_{jk}=I_r(E_j,x_k)$ is calculated for the potential
  $\Phi(s)$ at each point on the energy-angular momentum grid
  $[E_j,x_k\,L_c(E_j)]$, using equation (\ref{eq:action}). 

\item A BH of mass $M_\bullet$ is added at the center of the galaxy, thereby
  modifying the potential to  $\Phi^*(s)=\Phi(s)-M_\bullet/s$.

\item In this new potential we compute the angular momentum of
  circular orbits as a function of energy, $L_c^*(E^*)$.

\item A new energy-angular momentum grid 
  $[E^*_j,x_k\,L_c^*(E^*_j)]$ is defined, in which the new energies
  $E^*_j$ match the new potentials $\Phi^*_j=\Phi^*(s_j)$ at the
  radial grid points. The new action $I^*_{jk}=I_r^*(E^*_j,x_k)$ is
  computed on the new energy-angular momentum grid
  $[E^*_j,x_k\,L^*_c(E^*_j)]$.

\item Since actions are conserved, a star at $(E,x)$ evolves during
  the growth of the BH to $(E^*,x^*)$ where
  $I^*_r(E^*,x^*)=I_r(E,x)$, $x^*\,L_c^*(E^*)=x\,L_c(E)$. We solve for
  the original $(E,x)$ at each point on the new energy-angular
  momentum grid $[E^*_j,x_k\,L_c^*(E^*_j)]$ and then compute the new DF as
  $f^*(E^*,x^*L^*_c)=f[E(E^*,x^*L^*_c)]$.

\item Using the DF $f^*$ and potential $\Phi^*$ we compute the
  resulting density $\rho^{**}$ from equation (\ref{rhoel}) and the
  potential $\Phi^{**}$ from Poisson's equation (\ref{eq:potnorm}). We
  then replace $\rho^*$ and $\Phi^*$ by $\rho^{**}$ and $\Phi^{**}$ and
  return to step 6. 

\end{enumerate}

After about 20 iterations this algorithm converges, yielding a
DF for the stellar system from which quantities of interest
like density, surface brightness profile, and velocity dispersion
can be computed.

\section{Excess mass}

\label{sec:excess}
\begin{figure}[h]
\vspace*{1em}
\includegraphics[width=0.47\textwidth,height=0.38\textwidth]{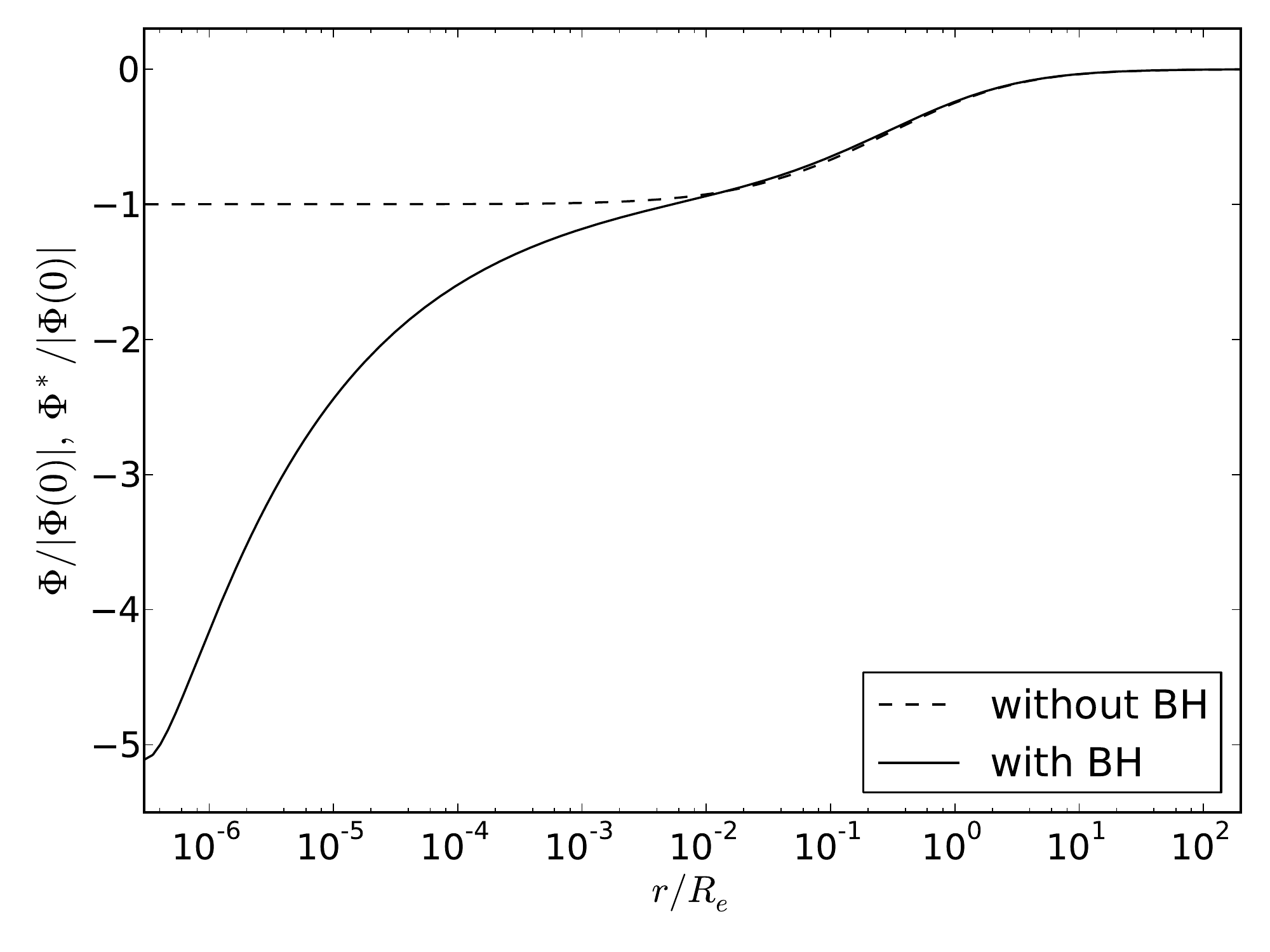}
\hspace*{-1em}
\includegraphics[width=0.48\textwidth,height=0.4\textwidth]{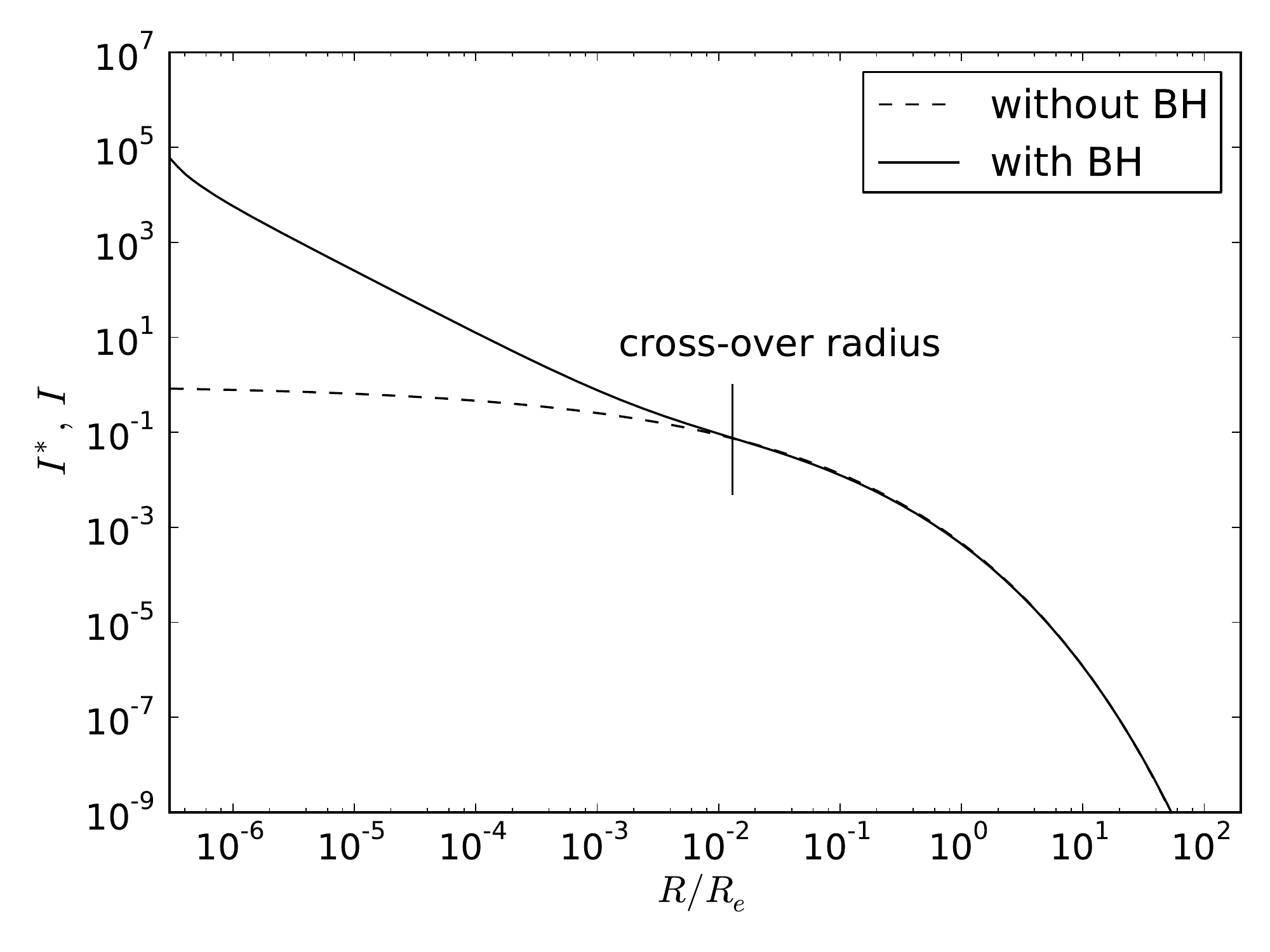}
\caption{\small Typical potential and surface density of a galaxy
  with and without a BH. The S\'ersic index $m=4$ and the black-hole
  mass relative to the galaxy mass is $M_\bullet/M_G=1.877\times 10^{-3}$.}
\label{fig:6}
\end{figure}

\noindent
We wish to quantify the excess light or stellar mass accumulated near
the center of the galaxy as the result of the adiabatic adjustment of
the stellar orbits to the growth of the BH. Since the excess has
appeared at the center only because of the rearrangement of the stars,
the total luminosity of the galaxy is unchanged: the density goes up
substantially close to the center, and decreases by a much smaller
amount at large radii. The definition of ``excess mass'' is therefore
somewhat arbitrary, but most reasonable definitions will give very
similar values. We define the``cross-over radius'' as the value
of the dimensionless radius $\eta=R/R_e$ at which the initial and final
surface-density profiles cross (see Figure \ref{fig:6}). Let $\Sigma(\eta)$ and $\Sigma^*(\eta)$ be the
dimensionless initial and final surface-density profiles, then $\eta_c$ is
defined by the equality $\Sigma^*(\eta_c)=\Sigma(\eta_c)$; \\
we have verified that
this solution exists and is unique. The excess mass is defined as
\begin{equation}   
M_e = \half\int^{\eta_c}_0 \eta [\Sigma^*(\eta)-\Sigma(\eta)]\,\ud \eta\,.
\label{excess}
\end{equation}

The factor $\half$ appears in the above equation so that the
dimensional definition of the excess mass is \mbox{$M_e=2\pi\int_0^{R_c}
[\Sigma^*(R)-\Sigma(R)]R\,\ud R$} (cf.\ eq.\ \ref{eq:potnorm})---in other
words it arises because the volume element is $4\pi s^2\,\ud s$ but the
area element is $2\pi \eta\, \ud \eta$. 

\section{Results and Discussion}

\label{sec:discuss}

\begin{widetext}
\begin{figure*}
\hspace*{-0.5cm}
\begin{minipage}{0.52\textwidth}
\includegraphics[height=2.3\textwidth,width=\textwidth]{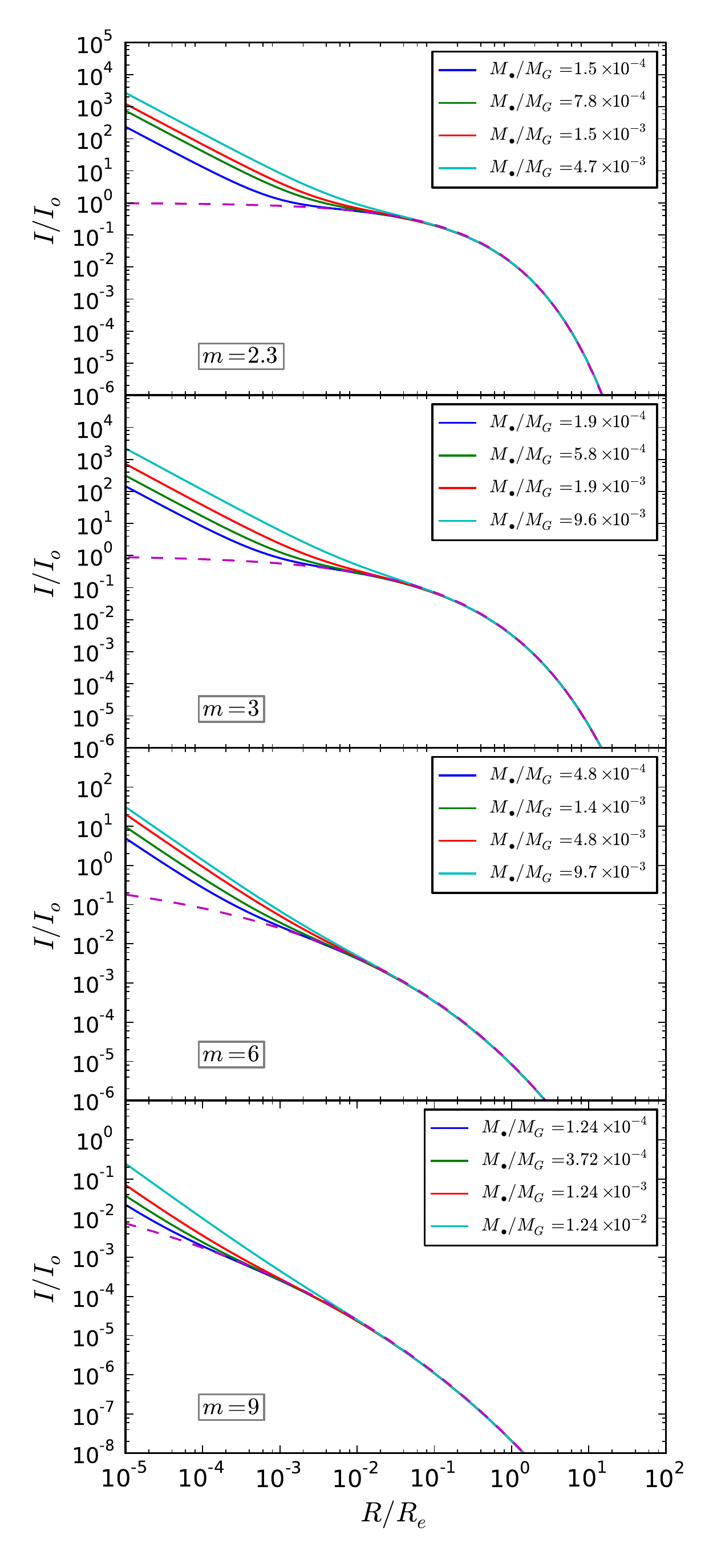}
\end{minipage}
\hspace*{-0.5cm}
\begin{minipage}{0.52\textwidth}
\includegraphics[height=2.3\textwidth,width=\textwidth]{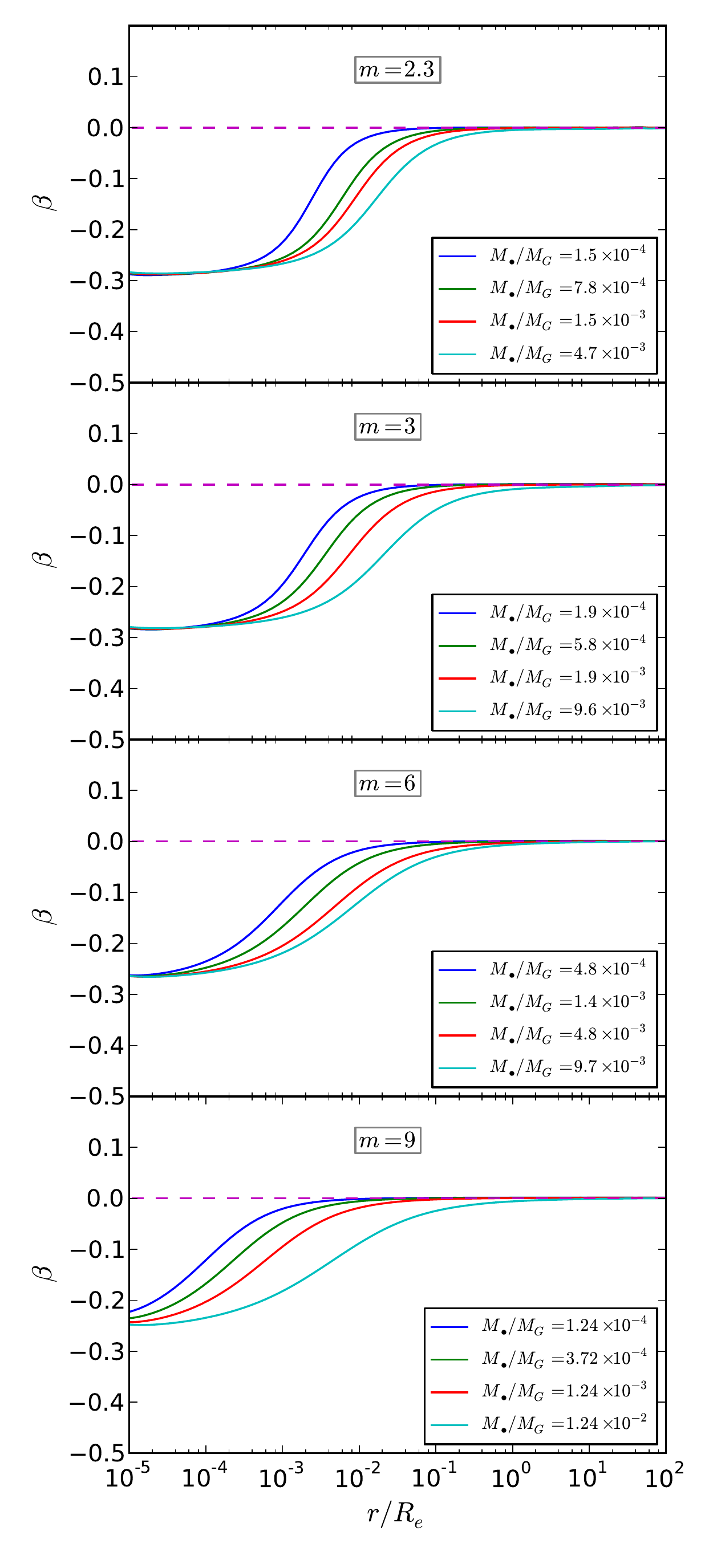}
\end{minipage}
\caption{\small Surface-brightness profile (left) and anisotropy
  parameter (right) after the adiabatic growth of a BH in a galaxy
  with S\'ersic index $m=2.3$, $m=3$, $m=6$, and $m=9$.}
\label{fig:8}
\end{figure*}
\end{widetext}

\noindent
The surface-brightness profile $I(R)$ and anisotropy parameter
$\beta(r)$ are plotted in Figure \ref{fig:8} for S\'ersic index $m=$
2.3, 3, 6 and 9 for dimensionless BH masses varying from $10^{-5}$ to
$5\times10^{-3}$. The BH masses are chosen such that they are
$10^{-4}$ to few times $10^{-2}$ of the galaxy mass ($M_G$).

\subsection{Surface-density cusp}

\noindent
The prominence of the surface-brightness or surface-density cusps
shown in Figure \ref{fig:8} varies with the S\'ersic index $m$. For
smaller $m$ the cusp is more prominent, partly because the initial
surface-brightness profile near the center is flatter. For large $m$,
even with the largest BH masses, the cusp caused by the BH is hardly
noticeable (see $m=9$ case in Figure \ref{fig:8}).

One can derive the approximate behavior of the surface-density cusp
analytically \citep{qhs95,GS99} 
\begin{equation}
\rho^*\sim r^{-\Gamma}\,,\quad \Sigma^*\sim R^{1-\Gamma}\,, \quad \Gamma = 2+\frac{1}{4-\gamma}=\frac{7m+2}{3m+1}.
\end{equation}
where $\gamma=(m-1)/m$ is the exponent of the initial
  cusp (see eq.\  \ref{scaling}). For the S\'ersic index $m=2.3$, $\rho^*\sim r^{-2.29}$ and $\Sigma^*\sim R^{-1.29}$
and for $m=9$, $\rho^*\sim r^{-2.32}$ and $\Sigma^*\sim R^{-1.32}$. We
see that the exponent of the central surface-density cusp does not vary strongly with
S\'ersic index. Thus the cusp becomes less prominent as $m$ increases
and the initial profile becomes more cuspy.

\subsection{Anisotropy and velocity cusp}

\begin{figure}
\hspace*{-2em}
\includegraphics[height=1.2\textwidth,width=0.53\textwidth]{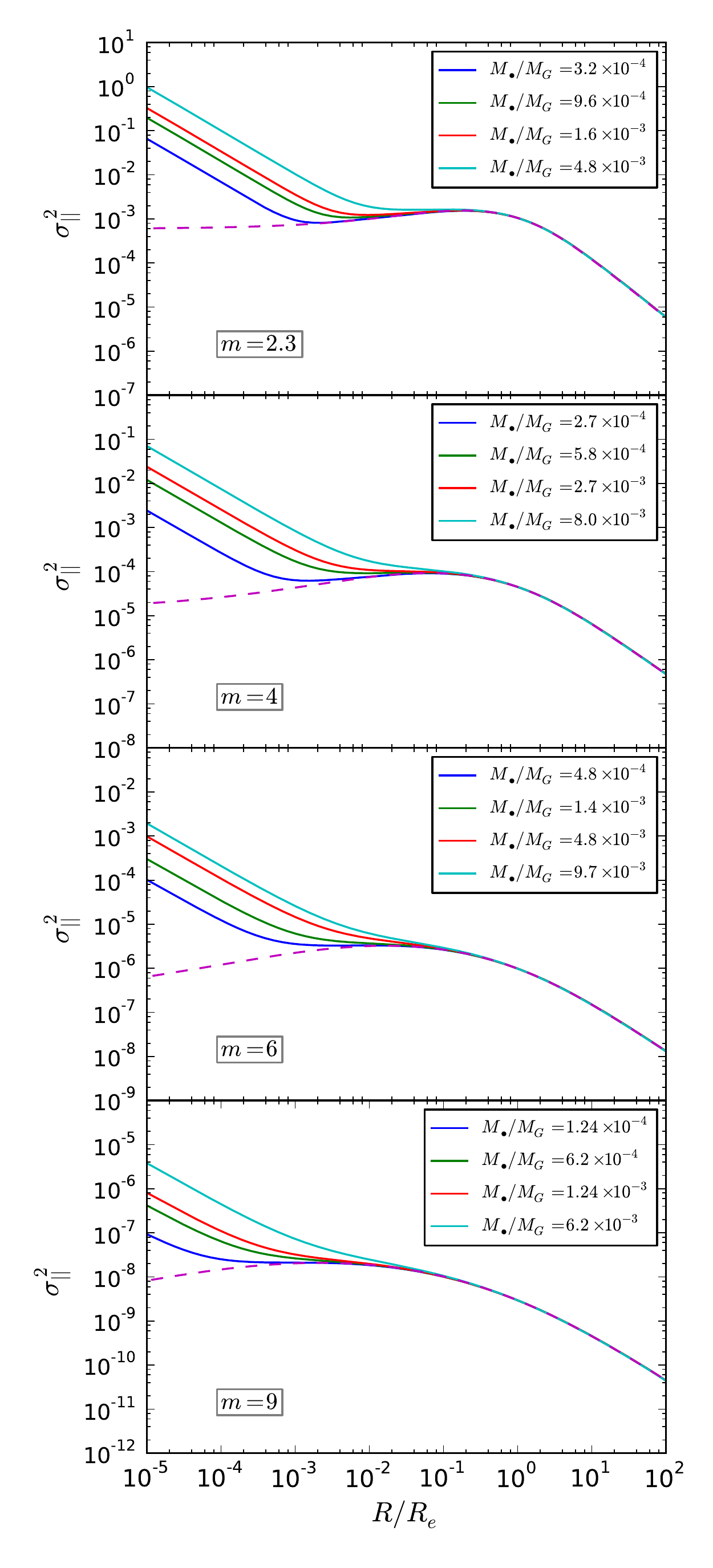}
\caption{\small Line-of-sight velocity dispersion after
  the adiabatic growth of a BH in a galaxy with S\'ersic index
  $m=2.3$, $m=4$, $m=6$, $m=9$. The dashed line is for the initial
  model without a BH.} 
\label{fig:9}
\end{figure}

\noindent
The line-of-sight velocity dispersion (eq.\ \ref{eq:los}) is shown in
Figure~\ref{fig:9}. Both the radial and tangential dispersions
$\overline{v_r^2}$ and $\overline{v_t^2}$ scale as $r^{-1}$ near the
BH and this scaling is evident in Figure~\ref{fig:9}. In contrast, the
velocity dispersion goes to zero at the center for all values of the
S\'ersic index $m$ for a galaxy without a central BH.

The anisotropy parameter $\beta$ is plotted for $m =2.3$ to
$m=9$ in Figure \ref{fig:8}. The plots show that as a result of the BH
growth the stellar DF develops a tangential
anisotropy $-\beta\simeq 0.24$--0.28 near the center. The anisotropy
depends only weakly on the S\'ersic index, and is independent of the
BH mass: as the BH mass grows the anisotropy simply extends to larger
distances. 

Using an approximate expression for the radial action near the center
of a S\'ersic model due to \cite{GS99}, we compute in the Appendix the
ratio of the radial and tangential velocity dispersions and thus the
anisotropy parameter $\beta$ (eq.\ \ref{eq:betadef}). When
sufficiently close to the center, the anisotropy ratio is independent
of the BH mass and the radius, and is determined solely by the initial
S\'ersic index $m$. The values of $\beta$ for $m=2.3$ and $m=9$ obtained
from equation (\ref{eq:anisot}) are $-0.313$ and $-0.306$,
respectively, and the corresponding values from Figure
\ref{fig:8} are $-0.285$ and $-0.247$. The constant tangential anisotropy near the center contrasts with the situation for initial galaxies having analytical cores as
described in \cite{BG84} and \cite{qhs95}: with analytic cores the
initial DF is constant as $r\to 0$ so the ratio
$\overline{v^2_r}/\overline{v^2_t}$ tends to $\half$ and $\beta$ tends
to zero. 
 
\subsection{Excess mass}

\begin{figure}
\hspace*{-1em}
\includegraphics[width=0.48\textwidth]{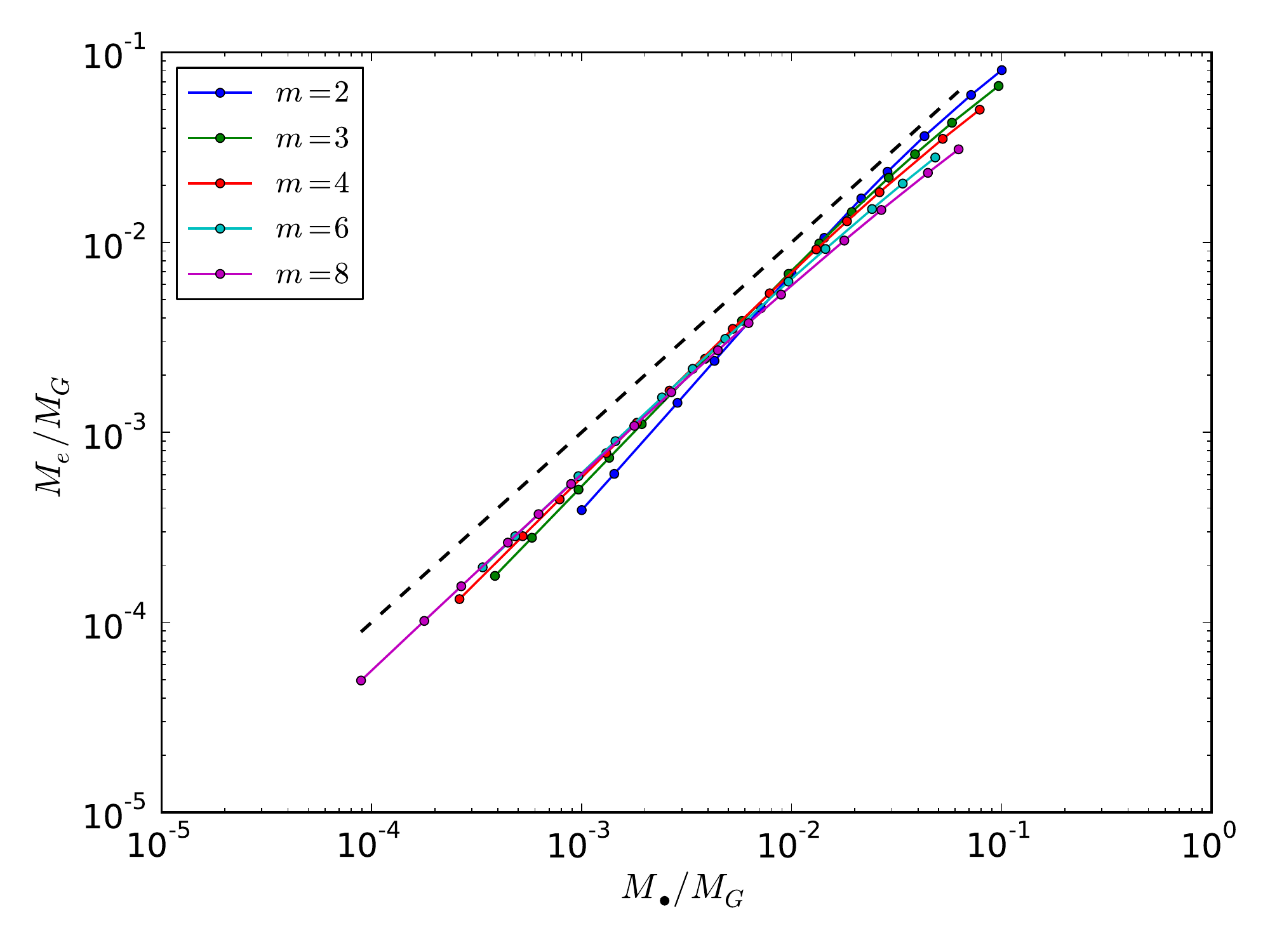}
\caption{\small Excess mass vs. BH mass, both scaled to the mass of
  the galaxy. The dashed line is $M_\bullet=M_e$.}
\label{fig:13}
\end{figure}
 
\begin{figure}
\hspace*{-1em}
\includegraphics[width=0.48\textwidth]{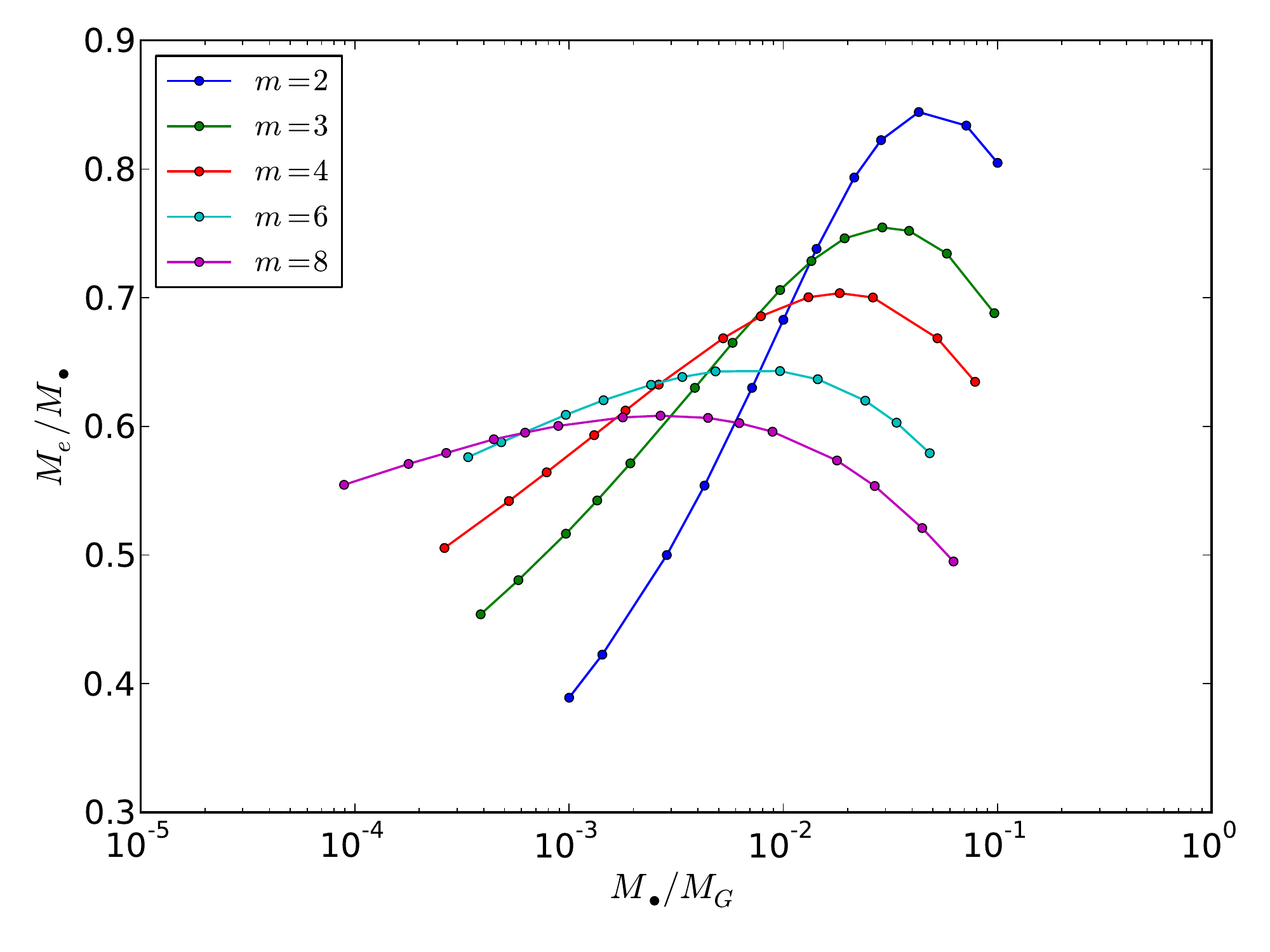}
\caption{\small Ratio of the excess mass to the BH mass, as a function
  of BH mass. The abscissa is plotted on a log scale.}
\label{fig:12a}
\end{figure}
\noindent
In Figure \ref{fig:13}, we plot the excess mass (as defined in
\S\ref{sec:excess}) with respect to the BH mass. We observe that the
excess mass scales approximately linearly with the BH mass, although
with some offset and curvature in the relation. To make these features
more apparent we plot in Figure \ref{fig:12a} the ratio of excess mass
to BH mass as a function of the BH mass.

We see that the excess mass $M_e$ is always smaller than the BH
mass $M_\bullet$ for all values of the BH mass and all values of the
S\'ersic index $m$.  There is a peak in the excess mass as a function
of BH mass for all values of the S\'ersic index, which occurs when the
mass of the BH is between about $10^{-3}$ and $10^{-1.5}$ times the
mass of the galaxy.  The maximum of the ratio $M_e/M_\bullet$
increases as $m$ decreases, consistent with the stronger visual
appearance of the cusp in Figure \ref{fig:8} at smaller S\'ersic
index, but this maximum varies by less than a factor of two over a
wide range of S\'ersic indices.

The dependence of the ratio of the excess mass to BH mass shown in
Figure \ref{fig:12a} can be fitted to a cubic polynomial, 
\begin{eqnarray}
y&=&\frac{M_e}{M_\bullet}=a\,\chi^3+b\,\chi^2+c\,\chi+d\; ;\nonumber \\
&\mbox{where}\quad &\chi = \log\left(\frac{M_\bullet}{M_G}\right)
\label{eq:cubic}
\end{eqnarray}

where the coefficients $a$, $b$, $c$ and $d$ are functions of $m$
alone. Table \ref{tab2} gives the fitted values of the coefficients of
the cubic polynomial; the ratio of black hole mass to the galaxy mass
at the maximum of $M_e/M_\bullet$ ($M_\bullet^{\rm max}/M_G $); and the maximum deviation of the fitting function $y_{\rm fit}$ from the numerically calculated values $y_{\rm num}$, that is, max($|y_{\rm fit}-y_{\rm num }|$).

\section{Comparison to observations}

\label{sec:obs}
\noindent
\begin{figure}[h]
\includegraphics[width=0.48\textwidth]{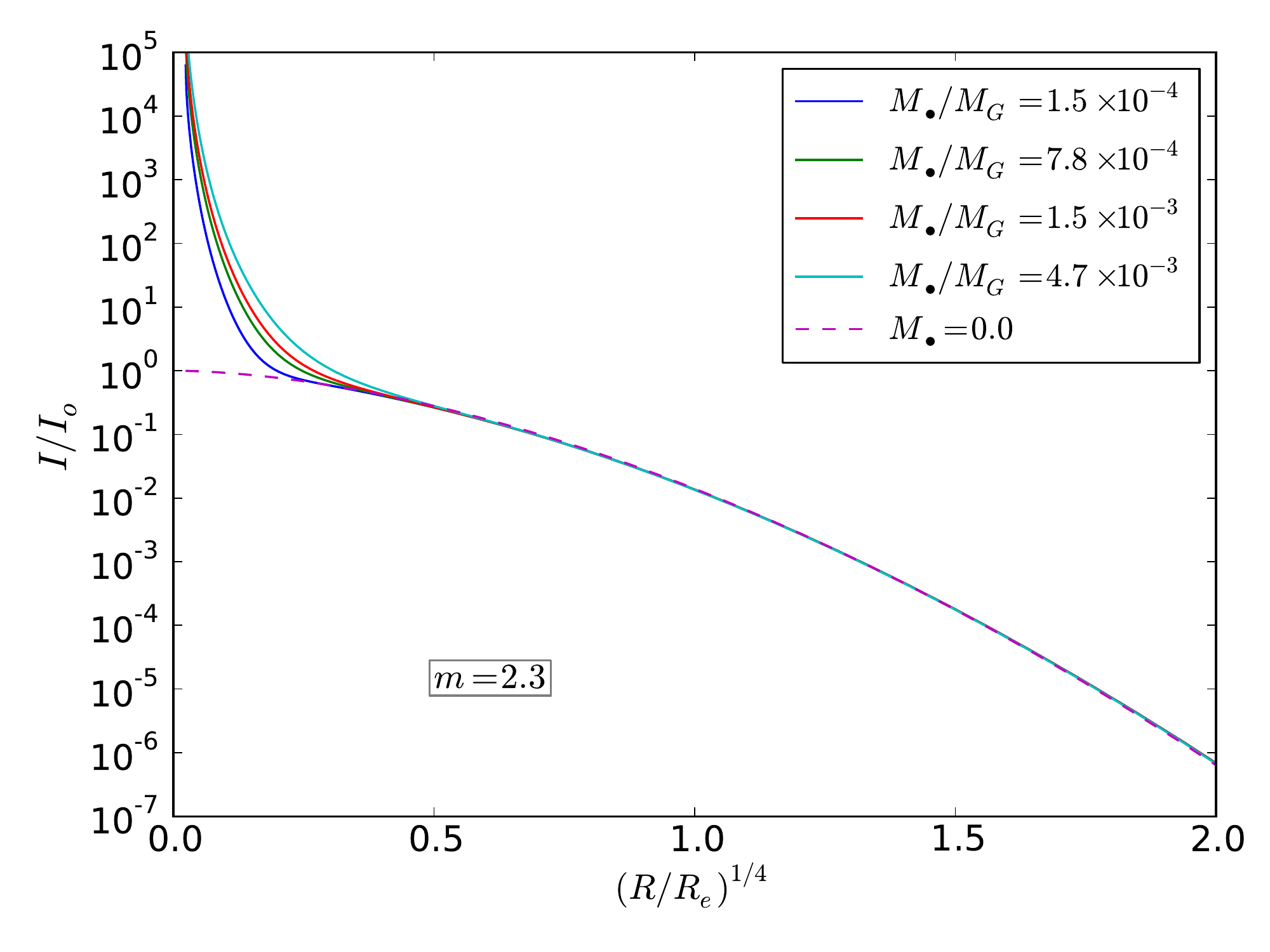}
\caption{Surface-brightness profile of the galaxy with S\'ersic
  index $m=2.3$, plotted with respect to $(R/R_e)^{1/4}$. The dashed line is $M_\bullet=0$.}
\label{fig:15}
\end{figure}
We now ask whether our results are consistent with observations of
excess light in elliptical galaxies. \cite{KOR09} estimate that the
excess light is 0.3\%--13\% of the total light in the galaxy (median
2.3\%); for comparison the typical ratio of BH mass to galaxy mass (in
stars) is $M_\bullet/M_G\simeq0.5\%$ \citep{kor13}. Figure
\ref{fig:12a} then implies $M_e/M_\bullet\simeq 0.7$ so $M_e/M_G\simeq
0.3\%$--0.4\%, at the lower end of the observed range. This result is
consistent with the top panel of Figure 42 of \cite{KOR09}, which
shows that $M_e/M_\bullet$ ranges between 1 and 50, with typical
values around 10. In other words the adiabatic model correctly
predicts that the excess mass should be roughly proportional to the
galaxy or BH mass, but the predicted magnitude is too small by a
factor 5--10. We conclude that the excess light in ellipticals does
not arise from adiabatic growth of a central BH. This conclusion is
consistent with the observation \citep{KOR09} that the excess light
often has disky isophotes, suggesting that it formed via dissipative processes rather
than by adiabatic ones---most likely late star formation due to gas
that has collected in the central region of the galaxy \citep[e.g.,][]{hop09}.
 
 Nuclei or nuclear star clusters are a different form of excess light
at the centers of galaxies. Kormendy et al. (Section 9.7) make a
distinction between the ``excess light'' in elliptical galaxies and
the ``nuclei'' in spheroidal galaxies. In Figure \ref{fig:15} the
surface brightness profile for a galaxy with S\'ersic index $m=2.3$ is
plotted using the same $(R/R_e)^{1/4}$ abscissa as in
Kormendy et al. Comparing this plot to Figures 25, 26, and 28 of
that paper, we notice that the slope of the surface-brightness
profile associated with nuclei in spheroidal galaxies, which is much
steeper than the profile associated with excess light in ellipticals,
is similar to what we obtain in the adiabatic-growth model. Moreover,
the median light fraction of nuclei in spheroidal galaxies is 0.3\%
\citep{KOR09}, much smaller than the fractional excess light in
ellipticals and consistent with our model if the nuclei contain
typical BHs with mass $M_\bullet/M_G\simeq0.5\%$ \citep{kor13}.

The observational relation between BHs and nuclear star clusters
remains unclear. \cite{cot06} and \cite{wh06} have argued that nuclear
star clusters and central BHs define a single smooth relationship
between mass (of the cluster or BH) and galaxy luminosity, with
clusters dominating at low luminosities and BHs at high
luminosities. \cite{KOR09} argue that this correlation is accidental,
but our understanding of the correlation between BH masses and their
host galaxy properties, particularly in low-luminosity and late-type
galaxies, is too limited for a definite conclusion. 

One attractive alternative to the formation of nuclear star clusters
by the adiabatic growth of a BH is that the clusters form
through in-spiral of globular clusters \citep{tos75,ant13,got14}. In
this case the BH could form before or after the bulk of the cluster
formation. 

\begin{widetext}
\center
\begin{table}[H]
\caption{The fitted values of the coefficients $a$, $b$, $c$ $d$ in
  equation (\ref{eq:cubic}) and {\ssc$M_\bullet^{\rm max}/M_G$}, the BH
mass at which $M_e/M_\bullet$ is largest.}
\label{tab2}
\begin{center}
\begin{tabular}{ccccccccccc}
\hline\hline
\rule{0pt}{3.1ex} 
\rule[-1.2ex]{0pt}{0pt}
 m &
 $-a$ &
 $-b$ &
$-c$ &
 $d$ &
\ssc $M_\bullet^{\rm max}/M_G$ &
max($|y_{\rm fit}-y_{\rm num}|$) \\
\hline
\rule{0pt}{3.1ex} 
\rule[-1.2ex]{0pt}{0pt}
\hspace*{-1em}
2 &\quad 1.21$\times 10^{-2}$ &\quad 1.84$\times 10^{-1}$ &\quad 0.77 &\quad $-0.140$ &\quad $5.21\times 10^{-2}$ &\quad $6.34\times 10^{-3}$  \\
3 &\quad 5.25$\times 10^{-3}$ &\quad 9.33$\times 10^{-2}$ &\quad 0.46 &\quad 0.0598 &\quad $3.03 \times 10^{-2}$ &\quad $5.183\times 10^{-3}$\\
4 &\quad 3.23$\times 10^{-3}$ &\quad 6.36$\times 10^{-2}$ &\quad 0.356 &\quad 0.0897 &\quad $1.76 \times 10^{-2}$ &\quad $3.107\times 10^{-3}$ \\
5 &\quad 2.47$\times 10^{-3}$ &\quad 5.13$\times 10^{-2}$ &\quad 0.312 &\quad 0.0784 &\quad $1.09 \times 10^{-2}$ &\quad $1.115\times 10^{-3}$\\
6 &\quad 1.88$\times 10^{-3}$ &\quad 4.14$\times 10^{-2}$ &\quad 0.272 &\quad 0.0810 &\quad $6.78 \times 10^{-3}$ &\quad $0.568\times 10^{-3}$ \\
7 &\quad 1.39$\times 10^{-3}$ &\quad 3.31$\times 10^{-2}$ &\quad 0.236 &\quad 0.0934 &\quad $4.21 \times 10^{-3}$ &\quad $0.761\times 10^{-3}$\\
8 &\quad 1.03$\times 10^{-3}$ &\quad 2.64$\times 10^{-2}$ &\quad 0.205 &\quad 0.1070 &\quad $2.65 \times 10^{-3}$ &\quad $0.968\times 10^{-3}$\\
9 &\quad 9.19$\times 10^{-4}$ &\quad 2.41$\times 10^{-2}$ &\quad 0.194 &\quad 0.0970 &\quad $1.83 \times 10^{-3}$ &\quad $0.574\times 10^{-3}$\\
\hline
\end{tabular}
\end{center}
\end{table}
\end{widetext}

\section{Conclusions}

\label{sec:conc}
\noindent
We have studied the effect of the adiabatic growth of a central black
hole on the $R^{1/m}$ or S\'ersic model of spherical galaxies,
following the methods described by \cite{Y80}. The black hole induces
a surface-brightness cusp at the center of the galaxy, which can be
described as ``excess light'' or ``excess mass'' above the inward
extrapolation of the best-fit S\'ersic profile for the outer galaxy.
At the smallest radii the surface brightness is found to vary with
radius as $I(R)\sim R^{-1.28}$ to $I(R)\sim R^{-1.32}$ for $m$ varying
from 2 to 9. Increasing the BH mass increases the strength of the cusp
without changing the logarithmic slope near the center. 
The line-of-sight velocity dispersion is found to scale as $R^{-1/2}$, which is the
expected behavior in a Keplerian potential; the anisotropy parameter 
$\beta$ near the center is between $-0.24$ and $-0.28$ (tangential
anisotropy). 

We calculated the excess mass (defined as the mass interior to the
radius on the sky where the initial and final surface densities were
the same) for S\'ersic models with varying indices $m$. The excess
mass is generally between 0.4 and 0.85 times the black-hole mass.

If the typical black-hole mass is $\sim0.5\%$ of the stellar mass in
the galaxy, the excess mass produced by adiabatic contraction is $\sim
5$--10 times smaller than the excess mass determined from S\'ersic
fits to the photometry of elliptical galaxies, but similar to the
masses of nuclear star clusters in low-luminosity galaxies of all
Hubble types. If black holes form by slow accretion of gas then
adiabatic contraction may play an important role in determining the
properties of nuclear star clusters. 

\acknowledgments

This work was supported in part by NASA grant NNX14AM24G and NSF grant
AST-1406166. Naveen Jingade(NJ) thanks S.Sridhar(RRI) for many fruitful discussions on this work and guiding through this project. NJ also thanks Raman Research Institute for hosting, where this work was carried out. NJ acknowledges the financial support of CSIR, India.

\bibliography{ref}

\section{}
\appendix
\section{}
\label{appen}

\noindent
The goal of this Appendix is to determine the velocity anisotropy near
the center of the cusp formed by the adiabatic growth of a central
BH.

The density for a spherically symmetric S\'ersic profile of index $m$
can be approximated near the center as (eq.\ \ref{scaling}):
\begin{equation} 
{\rho}_m(s)=\rho_o\,s^{-\gamma}\quad\mbox{where}\quad \gamma = (m-1)/m \,,
\end{equation}
and the corresponding potential is 
\begin{equation} 
{\Phi}_m(s)-\Phi_m(0)=\frac{\rho_o}{(3-\gamma)(2-\gamma)}s^{2-\gamma}=\phi_o\,s^{2-\gamma}\,.
\end{equation}
The phase-space DF for the above density and potential pair is given
by equation (\ref{eq:fdiv}):
\begin{equation}
f_m(E)\sim \left[E-\Phi_m(0)\right]^{-\alpha},\qquad \alpha =
\frac{6-\gamma}{2(2-\gamma)}=\frac{5m+1}{2(m+1)}.
\end{equation} 

An approximate expression for the radial action \citep{GS99} accurate
to 8\% for all $m>1$ is
\begin{equation}
I_r(E,L) =
\frac{2\pi}{d}\bigg\{-\frac{L}{\lambda}+\sqrt{2\,\phi_o}\left[\frac{E-\Phi_m(0)}{\phi_o}\right]^{(4-\gamma)/(4-2\gamma)}\bigg\}\,,
\label{eq:gs}
\end{equation} 
where 
\[\lambda  = \frac{2^{1/(2-\gamma)}(2-\gamma)^{1/2}}{(4-\gamma)^{(4-\gamma)/(4-2\gamma)}};\qquad d = \pi \frac{(2-\gamma )}{{B}\left[1/(2-\gamma) ,\frac{3}{2}\right]}.\] 
Once the black hole grows adiabatically, the galaxy potential at the
center is approximately Keplerian. The radial action is now given by
$I_r^*(E^*,L)=2\pi\big(M_\bullet/\sqrt{-2\,E^*}-L\big)$. The final DF
$f^*(E^*,L)$ is obtained by solving $I_r(E,L)=I_r^*(E^*,L)$ for $E$ as
a function of $E^*$ and $L$. Using the equation $E^* =\half v^2-M_\bullet/s$
and $L = s\,v_t$ to eliminate $E^*$ and $L$ in favor of the total and
tangential speeds $v$ and $v_t$, we find that the dependence of the
final DF on velocity at a given radius can be written as 
\beq f^* \sim 
\left[\left(\frac{1}{\lambda d}-1\right)x\sin\psi+\frac{x_m^2}{2\sqrt{x_m^2-x^2}}\right]^{-\delta},
\eeq
where
$$
\delta=\frac{6-\gamma}{4-\gamma}\quad;\quad 
x = sv \quad; \quad x_m = \sqrt{2M_\bullet\,s}\quad; \quad v_t =v\sin\psi\,.
$$
The quantities $x$ and $x_m$ are defined following \cite{BG84}. 

Multiplying $f^*$ by $v_r^2$ and $v_t^2$ and integrating over all
velocities we obtain the ratio of velocity dispersions and the
anisotropy parameter $\beta$: 
\begin{equation}
\frac{\overline{v_r^2}}{\overline{v_t^2}}=\frac{1}{2(1-\beta)}=\frac{\displaystyle
  \int_0^1\ud
  y\,y^4\int_0^\pi\,\ud\psi\,\sin\psi\cos^2\psi\left[\displaystyle
    \left(\frac{1}{\lambda d}-1   \right)y\sin\psi+\frac{1}{2\sqrt{1-y^2}} \right]^{-\delta }
}{\displaystyle \int_0^1\ud y\,y^4\int_0^\pi\,\ud\psi\,\sin^3\psi\left[\displaystyle
    \left(\frac{1}{\lambda d}-1
    \right)y\sin\psi+\frac{1}{2\sqrt{1-y^2}} \right]^{-\delta }}\, ,
\label{eq:anisot}
\end{equation}
where $y=x/x_m$. The only approximation in deriving this result is the approximate
form for the radial action, equation (\ref{eq:gs}). The quadratures 
are straightforward to evaluate numerically. 

\end{document}